\documentclass[aps,prl,onecolumn]{revtex4}
\usepackage{psfrag,graphicx}

\begin{document}

\title{Quantum Shuttle Phenomena\\
in a Nanoelectromechanical Single-Electron Transistor}



\author{D. Fedorets}
\email[]{dima@fy.chalmers.se}
\author{L. Y. Gorelik}
\author{R. I. Shekhter}
\author{M. Jonson}


\affiliation{Department of Applied Physics, Chalmers University of
  Technology \\ and G\"oteborg University, SE-412 96 G\"oteborg,
  Sweden}



\date{\today}
\begin{abstract}
An analytical analysis of quantum shuttle phenomena in a
nanoelectromechanical single-electron transistor has been
performed in the realistic case, when the electron tunnelling
length is much greater than the amplitude of the zero point
oscillations of the central island. It is shown that when the
dissipation is below a certain threshold value, the vibrational
ground state of the central island is unstable. The steady-state
into which this instability develops is studied. It is found that
if the electric field ${\cal E}$ between the leads is much greater
than a characteristic value ${\cal E}_q$, the quasiclassical
shuttle picture is recovered, while if ${\cal E}\ll{\cal E}_q$ a
new quantum regime of shuttle vibrations occurs. We show that in
the latter regime small quantum fluctuations result in large (i.e.
finite in the limit $\hbar \rightarrow 0$) shuttle vibrations.
\end{abstract}


\maketitle



The field of nanoelectromechanics has grown rapidly during the
last few years \cite{Roukes01,Craighead00,Cleland03}. In
particular, a nanoelectromechanical single-electron transistor
(NEM-SET) has been attracting a lot of theoretical
\cite{Gorelik98,Nishiguchi01,Boese00,
Fedorets02a,Fedorets02b,Fedorets03,McCarthy03,Nord02,Armour02,Alexandrov03,
Zhu03,Novotny03,Mitra03,Aji03,Lu03,Flensberg03,Smirnov03} and
experimental \cite{Park00, Erbe01} attention. A NEM-SET is a
single electron transistor (SET) where the position of a central
island (a small metal particle or a single molecule) is not
rigidly fixed but can oscillate under the influence of an elastic
potential. In \cite{Gorelik98} it was shown that in the regime
where the island motion can be treated classically and the
electron tunneling can be described by the Pauli master equation,
a new phenomenon
--- a so-called shuttle instability occurs. When a large
enough bias voltage is applied between the leads, the island
oscillates with an increasing amplitude until it reaches a stable
limit cycle where it oscillates with some constant amplitude. The
key issue in \cite{Gorelik98} was that as the island moves along
the classical trajectory its charge $q(t)$ correlates with its
velocity $\dot{x}(t)$ in such a way that the time average $
\overline{q(t)\dot{x}(t)}$ is positive. This results in
accumulation of energy in the vibrational degree of freedom and in
the development of the shuttle instability (see the review
\cite{Shekhter02})

Further miniaturization of the NEM-SET device brings up quantum
effects. In a nanometer-size metal particle, the electron energy
level spacing is about $10$~K and the discreteness of the electron
energy spectrum can no longer be neglected even for temperatures
of a few kelvin. In this case the characteristic de Broglie wave
length associated with the island can still be much shorter than
the length scale of the spatial variations of the "mechanical"
potential. If so, the motion of the island can be treated
classically. Shuttle phenomena in this regime have been studied
theoretically in \cite{Fedorets02a}. However, the classical
analysis of the shuttle instability (performed in
\cite{Fedorets02a}) is limited to displacements that exceed the
amplitude $x_{0}\equiv\sqrt{\hbar/(M\omega)}$ of the zero point
oscillations of the island ($M$ is the mass of the island and
$\omega$ its vibration frequency). This quantum limitation raises
the question whether or not a threshold value exists for the
displacement in order for a shuttle instability to develop. To
answer this question a quantum theory of the shuttle instability
must be developed. Moreover, the quantization of the island motion
might also effect the steady-state regime that develops.

Different aspects of the NEM-SET in the regime of quantized
mechanical motion of the island have already been studied
\cite{Boese00,McCarthy03,Alexandrov03,Zhu03,Mitra03,Aji03,Lu03,Flensberg03}.
However, no shuttle instability was found because either the
coupling between electron tunnelling and mechanical vibrations was
ignored \cite{Boese00} or strong dephasing in the mechanical
dynamics was expected
\cite{Alexandrov03,Zhu03,Mitra03,Aji03,Lu03,Flensberg03}. In this
Letter we will study the quantum dynamics of a NEM-SET for
arbitrary dissipation rates. A study complementary to ours has
recently been carried out by Novotny {\em et al}. (compare
\cite{Fedorets02b} and \cite{Novotny03}).  However, the numerical
analysis reported in Ref.~\onlinecite{Novotny03} was done only for
the case when a relatively small number of excited vibrational
states are involved. This is the case only if the amplitude of
zero point oscillations $x_{0}$ is of the order of the electronic
tunnelling length $\lambda_{\star}$ and if the dissipation is
large enough. Here we present a complementary analytical study
valid under the more realistic condition,
$x_{0}/\lambda_{\star}\equiv\lambda^{-1}\ll 1$.

We will formulate the problem at hand in terms of the
dimensionless displacement $x\equiv X/x_0$ and momentum $p\equiv
x_0 P/\hbar$ of the island. If we measure all lengths in units of
$x_0$ and all energies in units of $\hbar \omega$, the Hamiltonian
of the system reads
\begin{eqnarray}
    H = \sum_{\alpha, k} \epsilon_{\alpha k} a_{\alpha k}^{\dag}
a_{\alpha k}+  \left[{\epsilon}_0- dx\right] c^{\dag} c + H_{osc}
+ \sum_{\alpha, k} {T}_{\alpha}(x)\left[ a_{\alpha k}^{\dag} c  +
h.c. \right] + H_B + H_{B-osc}\,, \label{Hamiltonian}
\end{eqnarray}
where  $a_{\alpha k}^{\dag}$ creates an electron with momentum $k$
in the corresponding lead, $\alpha = L, R$ is the lead index,
$c^{\dag}$ creates  an electron on the single energy level in the
island, $d \equiv e{\cal E}/(M\omega^2 x_0)$ is the shift in the
equilibrium position of the oscillator due to the electric field
${\cal E}$  between the leads, $H_{osc}\equiv \left[{p^2} +
{x^2}\right]/2$ is the free oscillator Hamiltonian and
${T}_{L,R}(x) = T_{L,R}(0) \exp[\mp x/\lambda]$. We assume that
the electrons in each electrode are non-interacting with a
constant density of states ${\cal D}_\alpha$ and that all relevant
energies are small compared to the level spacing in the central
island which for typical systems under consideration exceeds
$100$~meV. In this case only one single level in the island is
relevant to the problem. The term $H_B$ describes a heat bath and
the last term $H_{B-osc}$ relates to the coupling between the
oscillator and the bath \cite{Weiss99}. We assume that this
coupling is linear in $x$ and treat it in weak-coupling limit. For
simplicity we will consider only zero temperature case.

The time-evolution of the system is governed by the Liouville-von
Neumann equation for the total density operator. After projecting
out the leads and the thermal bath we obtain an equation of motion
(EOM) for the reduced density operator $\rho$ of the vibrational
degree of freedom and the electronic state in the island. Under
conditions of large bias ($eV\gg \hbar \omega, \epsilon_0$), the
EOM for $\rho$ becomes Markovian (for details see
\cite{Fedorets02b,Novotny03, Mozyrsky02,Gurvitz}):
\begin{eqnarray}\label{master}
\partial_t \rho =   -i\left[H_v+(\epsilon_0 - d x)
c^{\dag}c , \rho \right] + \pi {\cal D}_L
      \left( 2 \hat{T}_{L} c^{\dag} \rho c
      \hat{T}_{L}  -
      \left\{ \hat{T}_{L}^2  c c^{\dag}, \rho \right\}
       \right)
     + \pi {\cal D}_R \left( 2\hat{T}_{R}^{\dag}
c \rho c^{\dag} \hat{T}_{R}
 - \left\{ \hat{T}_{R}^2
c^{\dag}  c , \rho \right\}  \right) + {\cal L}_{\gamma} \rho\,,
\end{eqnarray}
where ${\cal L}_{\gamma} \rho \equiv
-\frac{i\gamma}{2}\left[x,\left\{p,\rho\right\}\right] -
\frac{\gamma}{2}  \left[x,\left[x,\rho\right]\right]$,
$\{\bullet,\bullet\}$ denotes the anticommutator, $\gamma\ll 1$ is
a dissipation rate and time is measured in units of $\omega^{-1}$.
It follows from Eq.~(\ref{master}) that the time-evolution of the
electronic off-diagonal elements of the reduced density operator
is decoupled from the evolution of the diagonal elements. After
shifting the origin of the $x$-axis to the point $x=d/2$ and
introducing $\Gamma_{\alpha}(x) \equiv 2\pi {\cal
  D}_\alpha T_\alpha^2(x+d/2)$ we get the system of
  EOMs for the diagonal elements $\rho_{0 0}\equiv <0|\rho|0>$ and
  $\rho_{1 1}\equiv <1|\rho|1>$, where $|1>=c^{\dag}|0>$:
\begin{eqnarray}
\partial_t \rho_{0 0} = -i\left[H_{osc}+\frac{d}{2} x,\rho_{0 0}\right]
-\frac{1}{2}
      \left\{ {\Gamma}_L(x), \rho_{0 0} \right\}
     + \sqrt{{\Gamma}_R(x)}\, \rho_{1 1}
     \sqrt{{\Gamma}_R(x)} + {\cal L}_{\gamma} \rho_{0 0}
\,,\label{R:0}\\
 \partial_t \rho_{1 1}  =
 -i\left[H_{osc}-\frac{d}{2} x, \rho_{1 1}\right] - \frac{1}{2}
  \left\{{\Gamma}_R(x),
\rho_{1 1}\right\}  + \sqrt{{\Gamma}_L(x)}\, \rho_{0
0}\sqrt{{\Gamma}_L(x)} + {\cal L}_{\gamma} \rho_{1
1}\,.\label{R:1}
\end{eqnarray}
 In what follows we will deal with the
evolution of the density operator $\rho_{+}\equiv \rho_{0
0}+\rho_{1 1}$, which determines the expectation values of the
observables in vibrational space, and $\rho_{-}\equiv \rho_{0
0}-\rho_{1 1}$, which describes the shuttling of electrons.

The problem under consideration can be solved analytically in the
limit of a weak electromechanical coupling, $d/\lambda = e{\cal
E}/(M\omega^{2}\lambda_{\star}) \ll 1$.
 This
limit has already been used in the classical description of
shuttle phenomena \cite{Gorelik98,Fedorets02a}, where it assured
that an electric force acting on the charged island is much weaker
than a typical elastic force. For simplicity, we will also assume
that the tunnelling coupling is symmetric,
$\Gamma_{L}(0)=\Gamma_{R}(0)\equiv\Gamma/2$ and $\Gamma\ll 1$.

To study the vibrational dynamics near the ground state we use the
small parameter $\lambda^{-1}\ll 1$ to linearize the problem with
respect to the displacement $x$.  The linearized system of
equations, which describes the time-evolution of the expectation
value of the displacement $\bar{x}(t) \equiv \left<x\right>$ and
the momentum $\bar{p}(t)\equiv\left<p\right>$, of the island
($\left< \bullet\right> \equiv {\rm Tr}\{\rho_{+}(t)\bullet\}$)
has the following form
\begin{eqnarray}\label{Ehrenfest_1}
 \dot{\bar{x}} =  \bar{p}\,,\,
\dot{\bar{p}} = -\gamma \bar{p} -  \bar{x} - \frac{d}{2}\,
n_{-}\,,\,
 \dot{n}_{-} =   -
\Gamma n_{-} + \frac{2\Gamma}{\lambda}\bar{x}\,,
\end{eqnarray}
where $n_{-} \equiv 1 - 2{\rm Tr}\rho_{1 1}$. An analysis of
Eq.~(\ref{Ehrenfest_1}) shows that an initial deviation from the
equilibrium point grows exponentially in time with rate constant
$\alpha = (\gamma_{thr} - \gamma)/2$ if $\gamma < \gamma_{thr} =
\Gamma d/\lambda$. Therefore, when the dissipation is below the
threshold value $\gamma_{thr}$, the vibrational ground state
becomes unstable.

The exponential increase of the displacement drives the system
into the nonlinear regime of the vibrating dynamics, where the
system may reach a stable stationary state.
In order to study this regime we will use Wigner function analysis
suggested in \cite{Novotny03}. The Wigner distribution function
(WDF) corresponding to the density operator $\rho_{\pm}$ is
defined by
\begin{eqnarray}
  W_{\pm}(x,p) \equiv \frac{1}{2\pi} \int_{-\infty}^{+\infty} \, d \xi  e^{-ip\xi}
  \left<x+\frac{\xi}{2} \left| \rho_{\pm} \right|
  x-\frac{\xi}{2}\right>\,.
\end{eqnarray}
 After rescaling
the displacement, $X \equiv x/\lambda$, and momentum,
 $P \equiv p/\lambda$, we obtain
the following EOMs for the WDFs
\begin{eqnarray}
\partial_t{W}_{+} = \left[X\partial_P -P\partial_X +
    \hat{L}_1\right]W_{+}
+ \hat{L}_2 W_{-}\,, \\
\partial_t{W}_{-} = \left[X\partial_P - P\partial_X +
   \hat{L}_1  - {\Gamma}_{+}\right]W_{-}
 +\left[\hat{L}_2 + {\Gamma}_{-}\right]W_{+}\,,
\end{eqnarray}
where ${\Gamma}_{\pm}\equiv \Gamma_R(X)\pm \Gamma_L(X)$ and
\begin{eqnarray}
\label{L} \hat{L}_1 \equiv  {\gamma} \partial_P P +
\frac{\gamma}{2\lambda^2}\partial_P^2 -
    \frac{{\Gamma}_{+}}{2}
 \sum_{n=1}^{\infty} \frac{(-)^{n}}{\lambda^{4n}(2n)!}\partial_P^{2n}\,,
  \\
\hat{L}_2 \equiv
 \frac{d}{2\lambda}\partial_P + \frac{{\Gamma}_{-}}{2}
 \sum_{n=1}^{\infty}
 \frac{(-)^{n}}{\lambda^{4n}(2n)!}\partial_P^{2n}\,.
\end{eqnarray}
 It is convenient to study the steady-state solution in polar
coordinates, $X=A\sin\varphi$, $P=A\cos\varphi$. In these
coordinates, the steady-state solution is determined by the system
of equations
\begin{eqnarray}
\left[\partial_\varphi -\hat{L}_1\right] W_{+} = \hat{L}_2
W_{-}\,,
\label{StedyEOMforWplus:1}\\
\left[\partial_\varphi  + {\Gamma}_{+} -\hat{L}_1\right]W_{-}
    = \left[\hat{L}_2 + {\Gamma}_{-}
 \right]W_{+}\,, \label{StedyEOMforWminus:1}
\end{eqnarray}
with the periodic boundary conditions
$W_{\pm}(A,\varphi+2\pi)=W_{\pm}(A,\varphi)$. After eliminating
$W_{-}$ from the system of
 Eqs.~({\ref{StedyEOMforWplus:1}) and ({\ref{StedyEOMforWminus:1}), we
 get a closed equation for $W_{+}$
 \begin{eqnarray}
\left[\partial_\varphi - \hat{L}\right] W_{+}=0
\,,\label{StedyEOMforWplus:2}
\end{eqnarray}
where $\hat{L} \equiv \hat{L}_1 + \hat{L}_2 [1-\hat{G}_0
\hat{L}_1]^{-1}\hat{G}_0[{\Gamma}_{-}+\hat{L}_2]$ and $\hat{G}_0
\equiv \left[\partial_\varphi + {\Gamma}_{+}\right]^{-1}$ is
defined on the space of functions which are $2\pi$-periodic in the
variable $\varphi$. It is convenient to define a projector ${\cal
P}$ which maps a $2\pi$-periodic function $f(\varphi)$ to its
mean: ${\cal P}f(\varphi)\equiv \int_0^{2\pi} f(\varphi)
d\varphi/(2\pi)$ and a projector ${\cal Q}\equiv 1-{\cal P}$. We
use these projectors to decompose $W_{+}$ into two parts:
$W_{+}(A,\varphi) = \bar{W}_{+}(A) + \tilde{W}_{+}(A,\varphi)$,
where $\bar{W}_{+} \equiv {\cal P}W_{+}$ and  $\tilde{W}_{+}
\equiv {\cal Q}W_{+}$.
 By inserting this
decomposition into} Eq.~({\ref{StedyEOMforWplus:2}) and acting on
this equation from the left with ${\cal P}$ and ${\cal Q}$,
respectively, we obtain two coupled equations for $\bar{W}_{+}$
and $\tilde{W}_{+}$:
\begin{eqnarray}
{\cal P} \hat{L} \left[\bar{W}_{+} + \tilde{W}_{+}\right] = 0
\,, \label{W:bar:1}\\
\left[\partial_\varphi -{\cal Q} \hat{L}\right]\tilde{W}_{+} =
{\cal Q} \hat{L} \bar{W}_{+} \,.\label{W:tilde:1}
\end{eqnarray}
Formally solving Eq.~(\ref{W:tilde:1}) for $\tilde{W}_{+}$ and
substituting the result into Eq.~(\ref{W:bar:1}) gives a closed
equation for $\bar{W}_{+}(A)$,
\begin{eqnarray}\label{W:bar:2}
{\cal P} \hat{L} [1 -  \hat{g}_0
 {\cal Q} \hat{L} ]^{-1}\bar{W}_{+}(A) = 0\,,
\end{eqnarray}
where $\hat{g}_0 \equiv \partial_\varphi^{-1}$ acts in the space
of $2\pi$-periodic functions with zero mean. One can see from
Eq.~(\ref{W:tilde:1}) that $\tilde{W}_{+}$ is of lower order in
the small parameters $d/\lambda$, $\lambda^{-2}$ and $\gamma$ than
$\bar{W}_{+}$. Therefore, in the leading order approximation we
can write $W_{+}(A,\varphi)\approx \bar{W}_{+}(A)$. If we write
the LHS of Eq.~(\ref{W:bar:2}) in terms of $A$ and $\varphi$ and
expand it to second order in the parameters $d/\lambda$,
$\lambda^{-2}$ and $\gamma$, we get
\begin{eqnarray}\label{FP}
A^{-1}\partial_A A\left[ f(A) +
D(A)\,\partial_A\right]\bar{W}_{+}(A) = 0 \,,
\end{eqnarray}
where
\begin{eqnarray}\label{f}
f(A) \approx \frac{A}{2}\left[{\gamma}- \frac{d}{\lambda}
\alpha_0(A) -
\frac{1}{2\lambda^4}\alpha_1(A)\right]\,,\\
D(A) \approx \frac{\gamma}{4\lambda^2}+
\frac{1}{4\lambda^4}\beta_1 +
\left[\frac{d}{2\lambda}\right]^2\beta_2+\gamma\frac{d}{2\lambda}\beta_3>0\,,
\end{eqnarray}
$\alpha_0 \equiv -A^{-1}{\cal P}  \cos \varphi \, G$, $\alpha_1
\equiv A^{-1}{\cal P}  \cos \varphi \, {\Gamma}_{-}\,\partial_P
G$, $G\equiv \hat{G}_0 \Gamma_{-}$
 and $\beta_k = \beta_k(A,\Gamma)$.

We will see later that the functions $\alpha_0(A)$ and
$\alpha_0(A)$ determine the behavior of $\bar{W}_{+}$. The
positive function $\alpha_0(A)$ (see Fig.~\ref{alpha:0}) behaves
as $\Gamma(1+A^2/2)$ in the vicinity of $A=0$ and for large $A$ it
decreases as $\ln [2A/\Gamma]/(\pi A^2)$.  The function
$\alpha_1(A)$ (see Fig.~\ref{alpha:1}) is positive and grows as
$4A^2 \Gamma^3/9$ for small $A$. For large $A$ it is negative and
goes to zero as $-2/(\pi A^2)$.
\begin{figure}[htbp]
  \begin{center}
    \includegraphics[width = 7.5cm]{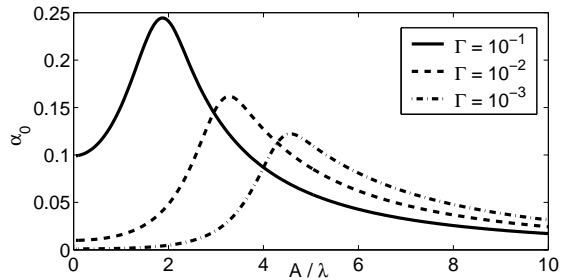}
    \vspace*{0cm}
    \caption{Function $\alpha_0(A)$ for $\Gamma/(\hbar\omega)=10^{-1}\,,\,10^{-2}\,,\,10^{-3}$
      }\label{alpha:0}
  \end{center}
\end{figure}
\begin{figure}[htbp]
  \begin{center}
    \includegraphics[width = 7.5cm]{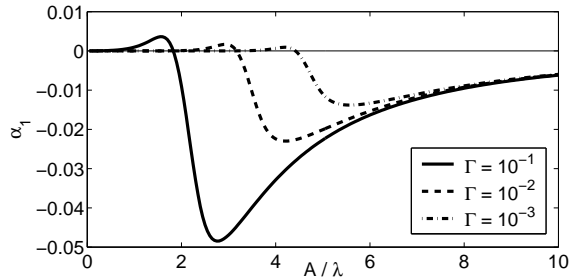}
    \vspace*{0cm}
    \caption{Function $\alpha_1(A)$ for $\Gamma/(\hbar\omega)=10^{-1}\,,\,10^{-2}\,,\,10^{-3}$
      }\label{alpha:1}
  \end{center}
\end{figure}

Solving Eq.~(\ref{FP}) gives
\begin{eqnarray} \label{W:solution}
\bar{W}_{+}(A) = {\cal Z}^{-1} \exp\left\{- \int_{0}^{A}\, d
A\,\frac{f(A)}{D(A)}\right\}\,,
\end{eqnarray}
where ${\cal Z}$ is a normalization constant determined by the
condition $2\pi\int_0^{\infty} dA\,A\, \bar{W}_{+}(A)=1$.

The steady-state solution $W_{+}$ is localized in the phase space
around points of maxima of $\bar{W}_{+}$. From
Eq.~(\ref{W:solution}) one can see that $\bar{W}_{+}$ has maxima
at points $A_{M}$, where $f(A_{M})=0$ and $f'(A_{M})>0$. In the
vicinity of these points, $\bar{W}_{+}$ is bell-shaped and can be
approximated by a Gaussian distribution function with variance
$\sigma^2\equiv D(A_{M})/f'(A_{M})$. Expanding Eq.~(\ref{f})
around $A=0$, we find that $f\sim A(\gamma-\gamma_{thr})/2$ as
$A\rightarrow 0$. Thus,
 $\bar{W}_{+}$ always has an extremum at $A=0$: a
maximum if $\gamma>\gamma_{thr}$ and a minimum if
$\gamma<\gamma_{thr}$. This reflects the fact that the vibrational
ground state is unstable when the dissipation is below the
threshold value (the shuttle instability discussed above).

The global behavior of $\bar{W}_{+}$ depends on the electric field
${\cal E}$. We have found two different regimes: an electric field
driven ``classical" regime, where ${\cal E}\gg {\cal E}_{q} \equiv
C(\Gamma) {\hbar^2}/{(eM\lambda_{\star}^3)}$
  and
a ``quantum" regime, when the electric field is weak, ${\cal E}\ll
{\cal E}_{q}$. The dimensionless $C(\Gamma) \approx
{\max[\alpha_1(A)] }/{ \max[\alpha_0(A)] }$ depends only weakly on
$\Gamma$ (for $\Gamma=10^{-3}\div 10^{-1}$, it is of the order of
$10^{-2}$).

In the classical regime, $\bar{W}_{+}$ has a maximum at finite
$A=A_{cl}$, if the dissipation is sufficiently weak, $\gamma
<\gamma_0 \equiv \max[\alpha_0(A)]\, d/\lambda>\gamma_{thr}$. The
width of the WDF around $A_{cl}$ is of the order of $\max\{
{d}/{\lambda}, \lambda^{-3}d^{-1}\} \ll 1$, which allows for a
classical interpretation of that regime. The value of $A_{cl}$
corresponds to the stable limit cycle amplitude of the classical
shuttle oscillations obtained in \cite{Gorelik98}. This amplitude
increases as the value of the dissipation decreases and since
$\alpha_0>0$, no stable state with finite $A_{cl}$ is possible
without external dissipation.

In the quantum regime, the structure of $\bar{W}_{+}$ is
determined by the quantum fluctuations of the island energy driven
by inelastic tunneling processes. In this case, the maximum of
$\bar{W}_{+}$ appears at finite $A=A_q$, when the dissipation is
below the critical value $\gamma_1 \equiv \max[\alpha_1(A)]\,x_0^4
/(2\lambda_{\star}^4)$. In contrast to the classical regime, $A_q$
at low dissipation  is determined by the zero of the function
$\alpha_1(A)$ (see Fig.~\ref{alpha:1}) and is still of the order
of one even when the dissipation is zero. Despite the fact that
the amplitude of the shuttle oscillations corresponding to the
maximum of WDF is much greater then the amplitude of the zero
point oscillations, the underlying steady state can not be
interpreted as classical because the width of $\bar{W}_{+}(A)$
around $A_q$ is no longer small compared to $A_q$.

The Wigner function ${W}_{+}$ describes the state of the
vibrational degree of freedom, while $W_{-}$ relates to the
correlations between the charge state of the island and the state
of the oscillator. It follows from
Eq.~({\ref{StedyEOMforWminus:1}) that  $W_{-}(A,\varphi) \approx
G(A,\varphi)\bar{W}_{+}(A)$. The WDF for the charged island is
given by $W_{1 1}\equiv[W_{+}-W_{-}]/2\approx
N(A,\varphi)\bar{W}_{+}(A)$, where $N\equiv \hat{G}_0 \Gamma_L$ is
the occupation of the classical shuttle with an amplitude $A$ and
frequency $\omega$ \cite{Fedorets02a}. The WDF for the charged
island exhibits qualitatively the same behavior as was observed
numerically in \cite {Novotny03}.

The steady-state current through the system is given by $I=e{\rm
Tr} [\Gamma_L(x) \rho_{0 0}] \equiv \frac{e}{2}\int\int dX dP\,
\Gamma_L(X)[W_{+}+W_{-}]$ (see \cite{Novotny03}). Using that
$W_{+}\approx \bar{W}_{+}$ and $W_{-} \approx G\bar{W}_{+}$, we
get
\begin{eqnarray}
I  \approx \left<I_{cl}(A)\right>\,,\quad
\left<\bullet\right>\equiv 2\pi\int_0^{\infty} dA \,A
\bar{W}_{+}(A) [\bullet]\,,
\end{eqnarray}
where $I_{cl}(A)\equiv e {\cal P}\Gamma_L(X)\hat{G}_0 \Gamma_R(X)$
is the time-averaged current through the classical shuttle
\cite{Fedorets02a}. Therefore, the current $I$ is given by the
classical expression $I_{cl}(A)$ averaged over the probability
distribution $\left<\bullet\right>$ of the amplitude $A$. The
function $I_{cl}(A)$ grows monotonically  from the tunneling
current $e\Gamma/4$ at $A=0$ to the shuttle current
$I_{shuttle}\equiv e\omega/(2\pi)$, which is reached already at
the amplitudes $A_{shuttle}\simeq 1$. In the classical regime the
distribution is narrow and $I \approx I_{cl}(A_{cl})$, where
$A_{cl}$ is the stable limit cycle amplitude of the classical
shuttle oscillations. Thus, in the classical regime the steady
state current is the same as in the classical shuttle case
\cite{Fedorets02a}. In the quantum regime, $A_q$ grows (as the
dissipation decreases) only as far as the zero $A_0$ of the
function $\alpha_1(A)$, but since $I_{cl}(A_0)\approx
I_{shuttle}$, one can see that the small quantum fluctuations
result in a large shuttle current even at ${\cal E}=0$ (as was
observed numerically in \cite {Novotny03}). This amplification is
another manifestation of the mechanical instability in a
non-equilibrium NEM-SET.

In conclusion, we have studied quantum shuttle phenomena in the
NEM-SET in the realistic limit, when the electron tunneling length
is much greater then the amplitude of the zero point oscillations
of the island. It is shown that when the dissipation is
sufficiently low, the vibrational ground state of the central
island is unstable. This shuttle instability develops into the
steady state corresponding to pronounced shuttle vibrations. For
large electrical fields between the leads this steady-state regime
can be interpreted in classical terms. At low field a new quantum
regime has been found.

The authors would like to thank A.-P. Jauho and T. Novotny for
valuable discussions. Financial support from the Swedish
Foundation for Strategic Research (DF, LYG, RIS) and the Swedish
Research Council (LYG,RIS) is gratefully acknowledged.

\end{document}